# Prism XR - A Curated Exhibition Experience in Virtual Reality with Peer Annotation Features and Virtual Guides for Art and Archaeology Classes

Huopu Zhang
Georgia Institute of Technology, hzhang931@gatech.edu

*Abstract* - The Prism XR project is a curated exhibition experience in virtual reality (VR) for art and archaeology education with features designed for the enhancement of interactivity and collaborative learning. The project integrates peer annotations and a virtual exhibition guide to augment educational experiences. The peer annotation features are intended for facilitating visitor critiques and comments pivotal in fostering a dialog between the curator and the audience and a dialogue between the visitors in art and archaeology education, which are demonstrated to have positive impacts on the learning motivations and learning outcomes. The virtual exhibition guide is intended to address the issue of isolation in the virtual exhibition space and to increase interactivity in the virtual curatorial experiences. The code for Unity3D implementation of the project can be found at: https://github.com/AnuitLecoidel/prismxr

## 1 INTRODUCTION

In recent years, Augmented Reality (AR) and Virtual Reality (VR) technologies have been increasingly deployed in Art and Archaeology Education for their positive impacts on improving students' learning motivations, the mechanisms and effectiveness of which have been studied and analyzed by Ángela Di Serio (2013), Shackelford (2019), and Cecotti (2024), based on the ARCS model proposed by Keller (1987). One particular type of AR/VR application commonly used in this context is applications with curatorial functionalities in the virtual learning environment, which not only gives the teacher the ability to curate an eclectic exhibition previously constrained by limitations in space and the availability of artworks for students (Ángela Di Serio et al., 2013) but also presents the students the opportunity to engage in hands-on curatorial experiences themselves. Such experiences are significant for art and archaeology learning as they cultivate dialogues between objects, dialogues between curatorial positions, and dialogues with the public, three important aspects of art education (Vella, 2018). Additionally, scholars, including Papasarantou et al. (2023), have demonstrated that the virtual gallery can help promote problem-based learning.

To introduce curated content and to integrate curatorial activities in the classroom, one can either take advantage of one of the AR/VR applications with curatorial functionalities readily usable, such as Artsteps and Sketchfab or, if equipped with more skills in software and programming, leverage tools such as Unity3D and its Vuforia SDK for more advanced development. When evaluating these tools used for curatorial activities in art and archaeology education, two of the important aspects to take into consideration are the development cost (Sarosa et al., 2019; Muñoz et al., 2020; Bäck et al., 2019) and the variety of content formats supported, which directly influences the richness and attractiveness of the experience (Mehmet et al., 2012; Grenfell, 2013). These aspects vary greatly between different applications. Take Artsteps and Sketchfab as examples, while Artsteps offers a seamless curatorial process integrated end-to-end on their platform, minimizing the development cost, it has much bigger restrictions in the formats of the content in that only 2D artworks are supported, whereas Sketchfab is more cumbersome in terms of development but offers support for 3D objects and viewing in AR. In addition, there are many shared strengths and weaknesses across these AR/VR applications. Some common strengths include good support for a wide range of AR/VR devices as well as traditional means of display when AR/VR devices are not available, and intuitive user experience and interface to add annotations. On the other hand, common limitations include the absence of viewer critique and comment functionalities, which are important in building up the dialogue between the curator and the public (Vella, 2018), and the issue of isolation in the learning environment, which limits the way of learning to autonomous learning (Martin-Gutierrez, 2014) and reduces the level of interactivity.

Thus, a specific problem in the future development in this field is to make enhancements to the curatorial experiences available in existing applications to address the lack of learning companions and the lack of viewer input for the curated content. At the same time, good practices and features in Artsteps and Sketchfab, such as cross-platform support and ease of use of the annotation functions should be acknowledged and carried over in future developments.

## 2 RELATED WORK

This section includes related work in applications for curation in art and archaeology education in three categories - AR/VR platform applications with curating capabilities; ad hoc development of AR/VR museum experience; and academic work related to curating art in AR/VR.

*I. AR/VR platform applications with curating capabilities*

Among AR/VR platform applications with curatorial functionalities, Artsteps and Sketchfab are two examples that have seen repeated deployment in the teaching of art and archaeology. Each of these two applications has its own strengths relative to the other. For example, Artsteps does not support 3D objects, which means its virtual exhibition space cannot host sculptures, installations, or archaeological artifacts, while Sketchfab can display a full array of all of these various types of content in addition to 2D content such as paintings, adding to the level of engagement (Muñoz et al., 2020) and attractiveness of its exhibitions (Ángela Di Serio et al., 2013). On the other hand, unlike the seamless curating process on Artsteps through an intuitive interface, the utilization of Sketchfab incurs a significant amount of development cost because the AR/VR experience on Sketchfab requires the exhibition space and the objects to be integrated into the same 3D model, meaning every modification of the exhibition inevitably involves editing of the experience in a 3D modeling software, such as Blender, outside Sketchfab.

Apart from the relative strengths and weaknesses mentioned above. Artsteps, Sketchfab, and some other AR/VR applications with curation capabilities share some common advantages and limitations. As for common advantages, both Artsteps and Sketchfab accommodate a wide range of platforms and both have well-developed support for annotations, which is an indispensable part of our traditional museum and gallery experiences (Reitstätter, 2021). In terms of common limitations, neither application supports comments and critiques from the viewer, meaning the annotation features currently available in these applications only allow a unidirectional flow of information from the curator to the audience, rendering the dialogue between the curator and the public missing from these experiences. Additionally, the AR/VR exhibition on Artstep or Sketchfab does not support a collective viewing experience, and the viewer is to be isolated in the virtual exhibition space, meaning these exhibitions only offer autonomous learning but not collaborative learning at the moment (Martin-Gutierrez, 2014). Moreover, Hasenbein et al. (2022) maintain that the presence of peers or learning companions in the virtual learning environment promotes learning outcomes.

*II. Ad hoc development of AR/VR museum experience*

Apart from using platform applications such as Artsteps and Sketchfab to curate virtual exhibitions, there are many experiments in cultivating AR/VR museum experience involving ad hoc application developments using different AR/VR development software and SDKs, examples of which being Unity3D and the Vuforia SDK, which are used in the development of projects such as ARThings (Lupascu et al., 2021) and ArkaeVision (Bozzelli et al., 2019). This way of curating virtual museum experiences requires significantly more development time compared to using platform applications yet allows more freedom in developing features not subject to the limitations of a particular platform. Therefore, in the context of incorporating AR/VR curation in course curricula, this approach has a high development cost but also presents learning experiences with high interactivity and novelty (Ungerer, 2016). Moreover, projects like ARThings and ArkaeVision provide us with insights into the positive impacts of the inclusion of interactive characters such as learning peers, experiential guides, and companions, as well as the inclusion of collaborative features in the virtual experiences:

- **ArkaeVision Project**: The ArkaeVision Project creates a multimodal virtual environment for cultural experiences tailored in a site-specific fashion. For each development, a team consisting of technologists, artists, actors, and historians is involved. Take the digital recreation of Hera II Temple of Paestum as an example, the project starts with serious archaeological research through collaboration between scholars and artists, before moving on to the modeling stage where 3D models are built based on 2D images and 3D performance content is created based on performing done by professional actors and actresses and encoded through motion capture technologies. The final stage of development includes scripting and programming using the Unity3D engine to bundle and build the experience into a VR application to be deployed on devices such as HTC Vive (Bozzelli et al., 2019, pp 8-13). The highlight of this experience is the interactive characters inside the virtual experience, which proved to be highly effective in keeping the user engaged despite being in an isolated virtual environment.
- **ARThings**: ARThings is a smart museum application created using the Unity3D engine and its Vuforia AR SDK. The project is deployed in the Art Museum of Brasov, Romania, where a User-Acceptance Test was carried out (Lupascu et al., 2021). The highlight of this project is the implementation of features enabling social feedback, realized through an interactive chat functionality and user comment functionalities. The researchers find that these social features are frequently used by the audience and a survey shows that the users prefer such features to traditional audio guides, validating the user acceptance of these interactive features. Moreover, the user input of comments and sentimental feedback facilitates the bi-directional information flow between the curator and the audience and allows the curator to engage with viewer critique within the same AR environment.

*III. Academic work related to curating art in AR/VR*

There are a few related works in academia focusing on the impacts of a few mixed-reality features and design decisions on the curated museum experience and study outcomes. Among these, Dallas (2004) systematically discusses the notion of the 'presence of visitors' in virtual exhibitions and analyzes the different degrees of presence, from showing visitor footprints to having virtual characters addressing the visitors in the second person, to manifesting the visitors with a form and name, and how these different ways of representation of the visitor presence engage the visitors with interactions and collaborations in the virtual space. Dallas also proposes scenarios where visitors can contribute to the virtual exhibitions through commentary features parallel to our real-world experiences, an example of which is the implementation of a virtual exhibition guestbook.

Other studies on the design and added value of a virtual exhibition guide include the UIUX design principles discussed by Hammady et al. (2019), including the spatial relation between the user and the guide, the combination of visual and auditory interactions, the relation of the guide with the virtual environment, etc. Another study by Sylaiou et al. (2019) investigates the credibility and visitor engagement of virtual museum guides realized in different types of avatars - a museum curator, a museum security guard, and another museum visitor, and provides interesting findings in the different emotional and sentimental involvement elicited by a narrator guide's perceived status distance from the visitor.

## 3 COMMON WEAKNESSES

Some common weaknesses are observed in existing virtual experiences created with platform applications such as Artsteps and Sketchfab, as well as in many ad hoc developments implemented using Unity. Among these, two types of limitations are of particular attention in this study based on considerations of user experience, user engagement, and interactivity.

*I. Lack of support for audience input*

A crucial component of curatorial activities is the dialogue with the public (Vella, 2018), which, from an art and archaeology class standpoint, is the dialogue built up through collaborative efforts and peer critiques. Currently, this dialogue is missing from experiences created using Artsteps and Sketchfab - neither application has features allowing viewers to leave comments or critiques on the individual art pieces, the layout of the exhibition, or the exhibition scheme as a whole. This dialogue is also missing from most of the ad hoc VR exhibition experiences developed using Unity with the exception of ARThings. The social feedback features in ARThings are well-received in the context of a public museum (Lupascu et al., 2021), which prompts an investigation into the values of such features in a classroom setting.

*II. Isolation in the learning environment*

Hasenbein et al. (2022) have demonstrated the positive effect of the presence of peers in the virtual learning environment through experiments. The lack of peers not only results in the lack of references or learning companions but also removes the opportunity for collaborative learning, which supplements autonomous learning in art education - a good virtual learning environment needs to be able to address both the autonomous and the collaborative aspects in the learning experience of art and archaeology (Martin-Gutierrez, 2014). As of now, neither Artsteps nor Sketchfab allows multiple users to be present in the same exhibition space and have a collective experience. To experience the curated content, one has to be alone in the space.

## 4 ENHANCED VR-CURATED EXPERIENCES: THE PRISM XR PROJECT

To address the common weaknesses mentioned in the previous section, the Prism XR project, a prototype of virtually curated exhibitions is proposed. The prototype features enhancements to existing VR exhibition curating tools mainly in the following two areas: 1. Enhancements in the annotation features to allow visitor annotation and to cultivate collaborative learning; 2. the introduction of a virtual exhibition guide. The development work is done in Unity3D and is built into a Meta Quest VR application.

*I. Virtual Exhibition Environment Overview*

The virtual exhibition environment is set up in Unity using mid to low-poly assets created by AK Studio Art (Asset link: https://assetstore.unity.com/packages/3d/environments/art-gallery-vol-10-260671). The exhibition environment is a square space with high clearance and is divided into two sub-spaces by a division wall in the middle. The exhibition space features 11 interactable artworks, all of which are sculptures, and the exhibition is titled 'Forms of Thought - Where the Timeless and the Modern Converge'. The locomotion system in this environment is set up through floor teleportation - the visitor can use the controllers to point to a location on the floor and hit the grip button to get teleported to that specific spot.

*II. Enhanced annotation features with audience input*

The main design idea behind the audience annotation features is to enable the bi-directional information flow between the curator and the audience so that the dialogue between the curator and the public and the dialogue between visitors can be constructed (Vella, 2018; Dallas, 2004).

Additionally, The annotations not only add necessary background information about the exhibition but also introduce additional interactivity (Tran et al., 2022):

- **Visitor Annotation View**: This view is triggered by interacting with any of the artworks in the exhibition environment. This view consists of 3 sections - On the top is the curator annotation section with preset information left by the curator, serving the same function as the labels we see in a museum in real life. Below the curator annotation section are two expandable sections initially hidden to ensure the conciseness of the view. The first of these two expandable sections is the visitor annotations section displaying comments, questions, and critiques left by earlier visitors, arranged in chronological order from the most recent to the oldest. The expandable section is for leaving a new comment. This section consists of two input fields - the guest name, and the comment content. The visitor can use the virtual keyboard to input new comments and submit and the new comments will be available for later visitors.
- **Annotation Summary View**: This is a UI that provides a consolidated view of all the annotations in the exhibition in the form of a scroller view with a number of expandable sections, each corresponding to an artwork and containing all the visitor comments on that art piece. The annotation summary view is triggered through interaction with the laptop placed on one of the display cases.
- **The virtual guestbook**: At the entrance of the exhibition space, there is a virtual guestbook object with which the visitor can interact. The interaction brings the visitor to a virtual guestbook UI where the user can leave their overall impression and comments of the exhibition as well as read the notes left by other people who have visited and left comments. This is a UI that provides a consolidated view of all the annotations in the exhibition in the form of a scroller view with a number of expandable sections, each corresponding to an artwork and containing all the visitor comments on that art piece. The annotation summary view is triggered through interaction with the laptop placed on one of the display cases.

*III. Virtual exhibition guide*

A virtual exhibition guide in the form of a curator is present in the exhibition space. The introduction of this virtual guide is to add to the richness of interactions in the virtual exhibition space as well as to address the issue of loneliness in space (Shackelford et al., 2018). The avatar guide introduced by Abate et al. (2011) with hidden dialogues and period-specific outfits is a good example that increases viewer engagement.

- **Avatar of the virtual guide**: The design of the virtual guide in this project is intended as a follow-up on the findings on the emotional and sentimental aspects found in Sylaiou et al. (2019). The primary avatar will take the form of the curator of the exhibition. Future developments can explore an avatar in the form of a fellow visitor.
- **Functions of the virtual guide**: The interaction with the virtual guide is dialogue-based, where the visitor is presented with various dialogue options. The main functions of the virtual guide are to provide the background story of this exhibition, give tips on the highlights of this exhibition, and help the visitor navigate the space. It is possible to ask the virtual guide for a map of the exhibition as well. This dialogue-based story system is implemented using the ink narrative scripting language developed by Inkle Studio. (Ink scripting language documentation page: https://www.inklestudios.com/ink/)
- **The virtual guestbook**: At the entrance of the exhibition space, there is a virtual guestbook object with which the visitor can interact. The interaction brings the visitor to a virtual guestbook UI where the user can leave their overall impression and comments of the exhibition as well as read the notes left by other people who have visited and left comments. This is a UI that provides a consolidated view of all the annotations in the exhibition in the form of a scroller view with a number of expandable sections, each corresponding to an artwork and containing all the visitor comments on that art piece. The annotation summary view is triggered through interaction with the laptop placed on one of the display cases.

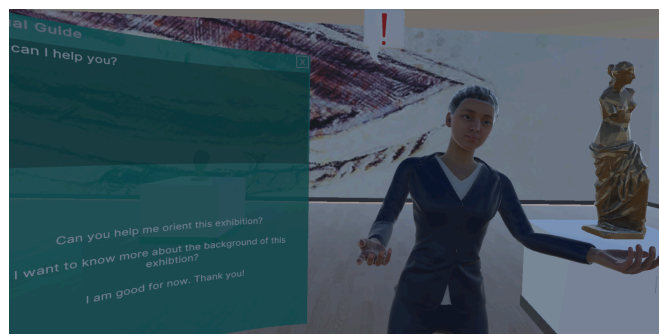

FIGURE I
DIALOGUE VIEW WITH THE VIRTUAL EXHIBITION GUIDE.

## 5 DISCUSSION

The peer annotation functions integrate audience input, and, along with such input from earlier visitors, traces of their presence, into the virtual exhibition experience while the virtual exhibition guide brings in a mobile avatar of the curator who created this exhibition and additional story-driven conversions. These functions are capable of enhancing classroom curatorial activities and fostering problem-based and collaborative learning. The richness of interactions consists of a wide range of activities, from reading other visitors' notes to leaving their own thoughts to conversations with the virtual exhibition guide, all of which contribute to a multi-layered interaction system in VR.

At the same time, some challenges commonly associated with VR environments remain evident. For one, the issue of dizziness shared by many similar VR experiences, which may detract from user engagement, can be present for some users and can hinder their taking advantage of either the peer annotation features or the virtual guide. Furthermore, there are concerns about the level of detail in the virtual exhibition. Although the exhibition was developed using assets of a similar level of resolution created by the same studio to maintain a plausible level of authenticity following the criteria of consistent resolution for authenticity outlined in Jacobson (2017), the fact that mid to low-poly assets are used throughout can cause a reduction in the level of immersiveness, which is crucial for a fully engaging virtual experience. Both of these two points need to be further studied and evaluated.

Additionally, while the annotation summary view can provide a standalone overview of user interactions with different artworks side by side, it currently offers the same user comment information as the individual artwork's annotation view. Improvements can be made in this view to add value to the learning experience. For example, enhancing the summary view to include features such as ranking of comments, ranking of artworks, and filters for comments could significantly augment its utility. Such enhancements would not only improve the curator's ability to consume and utilize user feedback effectively but also enrich the learning experience by providing more structured and accessible insights into the dialogue with the public (Vella, 2018).

## 6 DISCUSSION

User feedback is essential for the evaluation of VR experience app and as mentioned above, there are open questions on issues such as the level of immersion, the level of authenticity, the user experience, and the learning outcome, which need to be studied based on user feedback. For that, an evaluation plan is developed for work that still needs to be done. In short, user feedback will be gathered through a twofold process: firstly, through interviews with users after they have engaged with the full VR experience using a headset, and secondly, via a survey that includes questions on museum experiences, curatorial activities, and feedback on the 2D designs of the user interface.

*I. User interview based on in-game VR experience*

Firstly, questions on the ease of use of the application, which is fundamental to the effectiveness of technology (Bates et al., 2003), and the intuitiveness of the interactions with the objects in the space need to be asked. For example, for the interactions with the virtual exhibition guide, it is necessary to evaluate the dialogue options with the virtual guide for conciseness and effectiveness, focusing on the optimal amount of text and keyword emphasis to enhance user understanding without overwhelming them. For the peer annotation features, it will be helpful to assess the usability and impact of the note-leaving functionalities on art pieces, and in the virtual guestbook, examining its contribution to user engagement and learning outcomes can be meaningful. Furthermore, the overall user-friendliness of interactions within the virtual exhibition needs to be analyzed. Finally, it is necessary to make assessments on some common virtual reality issues such as dizziness and the smoothness of transitions, which can impact user engagement and thus, the efficacy of other features.

*II. Online User Survey*

Due to spatial constraints and constraints on the availability of devices, a full-blown VR experience might not always be available. To supplement this, twofold surveys are carried out to gather answers on museum VR experiences and curatorial activities and feedback on the UI design in the 2D format. The twofold surveys include one survey on museum VR experiences without specifically providing the context of classroom usage and a follow-up survey specifying the in-classroom usage.

One of the central questions in this study on virtually curated exhibitions is whether input, which includes critiques, comments, and other annotations, from the visitors, which in a classroom setting, includes the instructor and peers, helps enhance the learning experience in art education. This question can be asked both with and without specifically framing it in a classroom setting so that a comparison of the perceived value of such functions in a generic museum experience and in the background of an art or archaeology classroom can be made. Another question of concern is the role of the virtual guide in the exhibition space - how does an avatar of the curator compare with an avatar of a fellow visitor? Is the story-based dialogue system engaging and useful? Are the dialogue options concise and clear?

Lastly, an evaluation of the UI needs to be carried out. In general, the main concerns of the UI lie in its cleanness and conciseness. For example, in the annotation summary view and the visitor annotation view, there are multiple expandable sections in the same panel, potentially presenting too much information at the same time. Such arrangement of layouts needs to be assessed through user tests.